\documentclass[12pt]{article}
\usepackage[dvips]{graphics}
\usepackage{graphicx}
\usepackage[body={6.5in, 9.0in},left=1in, top = 1in]{geometry}
\usepackage{setspace}
\usepackage{varioref}

\newcommand{\p}{\partial}
\newcommand{\be}{\begin{equation}}
\newcommand{\ee}{\end{equation}}
\newcommand{\bea}{\begin{eqnarray}}
\newcommand{\eea}{\end{eqnarray}}
\newcommand{\bes}{\begin{equation*}}
\newcommand{\ees}{\end{equation*}}
\newcommand{\ba}{\begin{array}}
\newcommand{\ea}{\end{array}}

\begin{document}

\title{Comparison of analytical solution of spin-dependent DGLAP equations for $g_1^{NS}(x,t)$ at small $x$ by two methods}

\author{Neelakshi N K Borah $^{1}$\thanks{email:nishi\_indr@yahoo.co.in}, D K Choudhury $^{1,2,3}$, P K Sahariah $^{4}$\\ 
$^1$Department of Physics, Gauhati University, Guwahati-781014, India. \\
$^2$Physics Academy of North East, Assam, India.\\
$^3$Centre for Theoretical Studies, Pandu College,Guwahati-781012, India. \\
$^4$Department of Physics, Cotton College, Guwahati-781001, India. }

%\ead{nishi\_indr@yahoo.co.in}
\date{ }
\maketitle

\begin{abstract}
Analytical solutions for the non-singlet polarized parton distribution $\Delta q^{NS}(x,Q^2)=\Delta u(x,Q^2)-\Delta d(x,Q^2)$ are obtained by solving the DGLAP equation by two analytical methods: Lagrange's Method and Method of Characteristics. The realtive merits of the two methods are discussed while comparing with HERMES data in the small $x$ region. We also calculate the partial spin fractions carried by small $x$, non-singlet partons.\\	
keywords: {Deep inelastic scattering,Structure function, DGLAP equations.}\\
PACS no.:{12.38.-t;12.38.Bx;13.60.Hb;02.30.Jr}\\
\end{abstract}

\section{Introduction}
\subsection*{1.1 Spin structure function  }
The scale dependence of spin-dependent structure functions are in general interpreted by DGLAP \cite{D,G,L,A} equations. A lot of work has gone in understanding the nature of these structure functions over the past decades. Computation of exact \cite{AAC,bb,grsv,lss} as well as approximate analytical solutions \cite{6,7,8,9,10,11,12}
of these equations are equally important as one find simpler insight into the structure of the nucleon at least in some definite $x,Q^2$ range where the later is valid.\\
The aim of the present paper is exactly that, we will report a set of approximate analytical solutions of the LO DGLAP equation for the non-singlet polarized quark distribution $\Delta q^{NS}(x,Q^2)=(\Delta u+\Delta \bar{u})(x,Q^2)-(\Delta d+\Delta \bar{d})(x,Q^2)$. Based on the solution obtained by the Method of characteristics \cite{moc,moc1} and the Lagrange's method \cite{lagranges} we make a relative study of the two methods. While each method have been applied individually \cite{9,10,11,17,18,jksrb,jksrb1,nazir,mdv} earlier, the relative merit and demerit of the two has been reported only recently \cite{13} for unpolarized structure function  $F_2^{NS}(x,t)$. The present work reports a similar analysis for the polarized case, as was reported in ref \cite{12} briefly.\\
Our choice of studying $g_1^{NS}(x,t)$ is due to the reason of its simplicity: it evolves independently of singlet and gluon distribution and hence does not require solution of coupled DGLAP equation. The analytical solution of such coupled DGLAP equation invariably needs additional adhoc assumptions relating the two distributions not proved in QCD as noted in literature \cite{10,11,jksrb1,mdv}.
In our work we will show that even at LO level, analytical solutions have several interesting features, which needs attention.\\
The paper is organised as follows, in Section 2, we give the formalism, in Section 3, we present the results and discussions and in section 4 the conclusions are given.

\section{Formalism}
 \subsection*{2.1 Polarized Non-singlet DGLAP evolution equation  }
 The non-singlet polarized DGLAP equation in LO is,
\begin{equation}
\label{eqn:nonsingAP}
\frac{\partial}{\partial t} \Delta q^{NS}(x,t)=\frac{\alpha_s}{2\pi}\int_x^1 \frac{d z}{z}\Delta P_{qq}^{NS}(z)\Delta q^{NS}(\frac{x}{z},t) 
\end{equation}
Here $ t=\ln \frac{Q^2}{\Lambda^2}$ and  $\Delta P_{qq}^{NS}(z) $ is polarized splitting function. Introducing a variable $u=1-z$ and following the similar formalism as in ref \cite{13} with two different levels of approximation for small $x$, we can express Eq.(\ref{eqn:nonsingAP}) in two partial differential equations as, 
\begin{equation}
\label{eqn:pde1}
\frac{\partial \Delta q^{NS}(x,t)}{\partial t}=\frac{2}{\beta_0 t}\left[\frac{4}{3}\left(\log \frac{1}{x}+\frac{1}{2}\right) \Delta q^{NS}(x,t)+\frac{4}{3}\left(1-x-x \log \frac{1}{x}\right)\frac{\partial \Delta q^{NS}(x,t)}{\partial x}\right]
\end{equation} 
and
\begin{equation}
\label{eqn:pde2}
\frac{\partial \Delta q^{NS}(x,t)}{\partial t}=\frac{2}{\beta_0 t}\left[\frac{4}{3}\left(\log \frac{1}{x}+\frac{1}{2}\right) \Delta q^{NS}(x,t)+\frac{4}{3}\left(x \log \frac{1}{x}-x+x^2 \right)\frac{\partial \Delta q^{NS}(x,t)}{\partial x}\right]
\end{equation}
Eq.(\ref{eqn:pde1}) and Eq.(\ref{eqn:pde2}) are obtained from the same evolution equation Eq.(\ref{eqn:nonsingAP}) with two different levels of approximations given below by Eq.(\ref{eqn:series1}) and Eq.(\ref{eqn:series2}) respectively. While Eq.(\ref{eqn:pde1}) is based on the expansion given by,
\be
\label{eqn:series1}
\Delta q^{NS}(\frac{x}{z},t)=\Delta q^{NS}(x,t)+x\sum_{k=1}^\infty u^k \frac{\p \Delta q^{NS}(x,t)}{\p x}
\ee
Eq.(\ref{eqn:pde2}) considers only the first term of the expansion series on the RHS of Eq.(\ref{eqn:series2}),
\begin{equation}
\label{eqn:series2}
\Delta q^{NS}(\frac{x}{z},t)\approx \Delta q^{NS}(x,t)+xu \frac{\partial}{\partial x}\Delta q^{NS}(x,t)
\end{equation}

 We solve these two equations using two powerful methods for solving PDE: Method of Characteristics \cite{moc,moc1} and Lagrange's Method \cite{lagranges} .\\
To that end we express the Eq.(\ref{eqn:pde1}) and Eq.(\ref{eqn:pde2}) as,
\begin{equation}
\label{eqn:pdefrac}
Q(x,t)\frac{\partial \Delta q^{NS}(x,t)}{\partial t}+P(x,t)\frac{\partial \Delta q^{NS}(x,t)}{\partial x}=R(x,t)\Delta q^{NS}(x,t)
\end{equation}
with the forms of $Q(x,t)$, $P(x,t)$ and $R(x,t)$ being different for both the Eq.(\ref{eqn:pde1}) and Eq.(\ref{eqn:pde2}).
Where as,
\be 
Q(x,t)=t
\ee
\be
 P(x,t)=-\frac{2}{\beta_0} \left[\frac{4}{3}\left(1-x- x\log \frac{1}{x}\right)\right]
\ee
and
\be
 R(x,t,\Delta q^{NS}(x,t))=R'(x)\Delta q^{NS}(x,t)
\ee
 with
 \be 
 R'(x)=\frac{2}{\beta_0}\frac{4}{3}(\log \frac{1}{x}+\frac{1}{2})
 \ee
 for the Eq.(\ref{eqn:pde1}).\\
Similarly,
\be 
Q(x,t)=t
\ee
\be
 P(x,t)=-\frac{2}{\beta_0}\left[ \frac{4}{3}\left(x\log \frac{1}{x} -x+x^2\right)\right]
\ee
and 
\be  
R(x,t,\Delta q^{NS}(x,t))=R'(x)\Delta q^{NS}(x,t)
\ee 
 with 
\be  
 R'(x)=\frac{2}{\beta_0}\frac{4}{3}(\log \frac{1}{x}+\frac{1}{2})
\ee 
for the Eq.(\ref{eqn:pde2}). The only difference occurs in the structure of $P(x,t)$ ,Eq.(8) and Eq.(12).\\ 

\subsection*{2.2 Solution by the Method of Characteristics}
The method of characteristics \cite{moc,moc1} as a strong tool for solving a partial differential equation in two variables been well discussed in our recent work \cite{12,13}. In this formalism, the characteristic equations for the Eq(\ref{eqn:pde1}) and Eq.(\ref{eqn:pde2}) has the form:

\begin{equation}
\label{eqn:cheqn1}
\frac{d x}{d s}=P(x,t) 
\end{equation}
\begin{equation}
\label{eqn:cheqn2}
\frac{d t}{d s}=Q(x,t) 
\end{equation}
So, along the characteristic curve the PDE (Eq.(\ref{eqn:pde1})) and Eq.(\ref{eqn:pde2}) becomes an ODE:
\begin{equation}
\label{eqn:checurve}
\frac{d \Delta q^{NS}(s,\tau)}{d s}+c(s,\tau)\Delta q^{NS}(s,\tau)=0 
\end{equation}
Here,
\begin{equation}
c(s,\tau)=-\frac{2}{\beta_0} \frac{4}{3}\lbrace-\log\lbrace\tau\exp \left(\frac{t}{t_0}\right)^{\frac{8}{3\beta_0}}\rbrace+\frac{1}{2}\rbrace
\end{equation}
and it has the identical form in both the cases. The solutions of the characteristic equations Eq.(\ref{eqn:cheqn1}) and Eq.(\ref{eqn:cheqn2}) for Eq(\ref{eqn:pde1}) comes out as,

\begin{eqnarray}
\label{eqn:solcheqn1}
s=\ln\left(\frac{t}{t_0}\right)\\
\tau =x \exp{[(\frac{t}{t_0})^{8/3\beta_0}]}
\label{eqn:solcheqn2}
\end{eqnarray}
while for Eq.(\ref{eqn:pde2}), these are,
\begin{equation}
\label{eqn:solcheqn3}
s=\ln\left(\frac{t}{t_0}\right)
\end{equation}
\begin{equation}
\label{eqn:solcheqn4}
\tau' =x \exp{[-(\frac{t}{t_0})^{8/3\beta_0}]}
\end{equation}
The solutions Eq.(19-22) are in $(s,\tau)$ space. Using these solutions of the characteristic equations, we can express the solutions of the Eq.(\ref{eqn:pde1}) and Eq.(\ref{eqn:pde2}) in $(x,t)$ space in a more precise form as,\\
Eq.(\ref{eqn:pde1}), MOC1:

\begin{equation}
\label{eqn:solnspol1}
\Delta q^{NS}(x,t)=\Delta q^{NS}(x,t_0)\left(\frac{t}{t_0}\right)^{\tilde{n_1}(x,t)}
\end{equation}
where
\begin{equation}
\tilde{n_1}(x,t)=\frac{1}{\log \left( \frac{t}{t_0}\right)}\log \left(\frac{\Delta q^{NS}(\tau)}{\Delta q^{NS}(x,t_0)}\right)+\frac{\alpha_1}{\log \left( \frac{t}{t_0}\right)}
\end{equation}

Eq.(\ref{eqn:pde2})MOC2:

\begin{equation}
\label{eqn:solnspol2}
\Delta q^{NS}(x,t)=\Delta q^{NS}(x,t)\left(\frac{t}{t_0}\right)^{\tilde{n_2}(x,t)}
\end{equation}
where
\begin{equation}
\tilde{n_2}(x,t)=\frac{1}{\log \left( \frac{t}{t_0}\right)}\log \left(\frac{\Delta q^{NS}(\tau')}{\Delta q^{NS}(x,t_0)}\right)+\frac{\alpha_1}{\log \left( \frac{t}{t_0}\right)}
\end{equation}
with  
\begin{equation}
\alpha_1 =\frac{4}{3\beta_0}\left[1+2\log \frac{1}{x}\right]
\end{equation}

Eq.(\ref{eqn:solnspol1}) and Eq.(\ref{eqn:solnspol2}) are our solutions for $g_1^{NS}(x,t) $ obtained from the Eq.(\ref{eqn:pde1}) and Eq.(\ref{eqn:pde2}). The exponent $\tilde{n_1}(x,t)$ and $\tilde{n_2}(x,t)$ are different because $\tau$ and $\tau'$ as defined in Eq.(\ref{eqn:solcheqn2}) and Eq.(\ref{eqn:solcheqn4}) are not the same. We have assumed that $\log \frac{1}{x}\gg1$, as well as  $x \log \frac{1}{x}\ll1$ in deriving Eq.(\ref{eqn:solnspol1}) and Eq.(\ref{eqn:solnspol2}). Analytical solutions are possible only under these extreme conditions. Thus we see that both the solutions obtained from the same evolution equation with two different levels of approximation leads us to two results, with the difference in the solution of the characteristic equations only.\\

\subsection*{2.3 Solution by the Lagrange's Auxiliary Method}

To solve the equation Eq.(\ref{eqn:pde1}) by the Lagrange's Auxiliary Method\cite{lagranges}, we use the auxiliary system of equation given by Eq.(\ref{eqn:pdefrac}).\\
The general solution of the Eq.(\ref{eqn:pde1}) and Eq.(\ref{eqn:pde2}) are obtained by solving the following auxiliary system of ordinary differential equations,
\begin{equation}
\label{auxiliary}
\frac{dx}{P(x)}=\frac{dt}{Q(t)}=\frac{\Delta q^{NS}(x,t)}{R(x,t,\Delta q^{NS}(x,t))}
\end{equation}
If $u(x,t,\Delta q^{NS})=C_1$ and $v(x,t,\Delta q^{NS})=C_2$ are the two independent solutions of Eq.(\ref{auxiliary}),then in general, the solution of Eq.(\ref{eqn:pdefrac}) is
\begin{equation}
F(u,v)=0
\end{equation}
Where $F$ is an arbitrary function of $u$ and $v$. In our recent work we have applied Lagrange's method successsfully for the corresponding unpolarized structure function \cite{13}. Using the same approach with physically plausible boundary conditions, Eq.(\ref{eqn:pde1}) and Eq.(\ref{eqn:pde2}) give the specific solution for  $g_1^{NS}(x,t) $ as,\\
\be
\label{eqn:solnlagmethod1}
\Delta q^{NS}(x,t)=\Delta q^{NS}(x,t_0)\left(\frac{t}{t_0}\right)\frac{[X^{NS}(x)-X^{NS}(1)]}{[X^{NS}(x)-(\frac{t}{t_0})X^{NS}(1)]}
\ee
The explicit analytical form of $X^{NS}(x) $ in the leading $(\frac{1}{x})$ approximation for Eq.(\ref{eqn:pde1}) is,
\begin{equation}
\label{Xnsvalue1}
X^{NS}(x)=\exp[\frac{3\beta_0}{8}\log |\log \frac{1}{x}|]
\end{equation}
while for Eq.(\ref{eqn:pde2}), it is,
\begin{equation}
\label{Xns2}
X^{NS}(x)=\exp[\frac{3\beta_0}{8}\log(-1+\log \frac{1}{x})]
\end{equation}
At this level, the solutions Eq.(\ref{Xnsvalue1}) and Eq.(\ref{Xns2}) are distictively different as was the case with the method of characterisrics (Eq.(\ref{eqn:solnspol1}) and Eq.(\ref{eqn:solnspol2})). However as Eq.(\ref{Xns2}) is not real at $x=1$, it will not give a physically plausible solution of $\Delta q^{NS}(1,t)$ and can be reuled out on physical ground. But in the limit $\log \frac{1}{x}\gg1$, as has been used in the derivation of Eq.(\ref{eqn:solnspol1}) and Eq.(\ref{eqn:solnspol2}) for the method of characteristics,both Eq.(\ref{Xnsvalue1}) and Eq.(\ref{Xns2}) are identical. i.e.
In both the cases,
\begin{equation}
\label{xns1}
X^{NS}(1)\approx 0.
\end{equation}
Hence we get,
\begin{equation}
\label{solnlagmethod1}
\Delta q^{NS}(x,t)=\Delta q^{NS}(x,t_0)\left(\frac{t}{t_0}\right)
\end{equation}
as the analytical solution using Lagrange's method, for Eq.(\ref{eqn:pde1}) and Eq.(\ref{eqn:pde2}) at small $x$.
Unlike the solutions obtained using the method of characteristics Eq.(\ref{eqn:solnspol1}) and Eq.(\ref{eqn:solnspol2}), the solutions obtained by using the Lagrange's auxiliary method are same.\\ 
	In the next section we consider the phenomenological utility of our solutions with respect to each other vis-a-vis the available experimental data. We perform a $\chi^2$ test to test their compatibility with the data.\\

\section{Results and discussion}
\label{sec:results} 
We now compare our analytical solutions Eq.(\ref{eqn:solnspol1})(MOC1), Eq.(\ref{eqn:solnspol2})(MOC2) and Eq.(\ref{solnlagmethod1})(LM) with the HERMES \cite{HERMES} and COMPASS \cite{compass} data for the polarized non-singlet structure function $g_1^{NS}(x,t) $ related to the non-singlet polarized parton densites $\Delta q^{NS}(x,t)=\Delta u-\Delta d$, by using the relation \cite{HERMES},
\begin{equation}
\label{g1ns}
g_1^{NS}(x,t)=\frac{1}{2}\frac{1}{n_f}\sum_{i-1}^{n_f} e_i^{2} \Delta q^{NS}(x,Q^2)
\end{equation}
For any flavour, $n_f=2,3,4,5,6$, it turns out to be,
\begin{equation}
g_1^{NS}(x,t)=\frac{1}{9}\Delta q^{NS}(x,Q^2)
\end{equation}
The data of ref \cite{HERMES,compass} are available within the range $0.0264 \leq x \leq 0.7311$, $1.12GeV^2 \leq Q^2 \leq 14GeV^2$ for HERMES and $0.0046 \leq x \leq 0.566$, $1.1GeV^2 \leq Q^2 \leq 55GeV^2$ for COMPASS respectively. The approximate analytical solutions, although derived in the ultra small $x$ limit; ($\log \frac{1}{x}\gg1$, as well as  $x \log \frac{1}{x}\ll1$), we study if they are compatible with the available data at the range $x$($x\geq 0.0264$) and ($x\geq 0.0046$) reasonably \cite{saiful}.\\

We consider data for comparison from 36 and 14 individual kinematic bins from (HERMES) and (COMPASS) respectively as well as statistical uncertainities for each bin. To evolve our solutions we take the input distribution from LSS05\cite{lss} and consider $Q_0^2=1GeV^2$ and in LO $\Lambda^2=0.393GeV^2$ \cite{PDG}. \\

To derive the final $g_1^{NS}(x,t)$, the follwing relation is used,\cite{HERMES}
\begin{equation}
g_1^{NS}\equiv g_1^P-g_1^n=2\left[g_1^P-\frac{g_1^d}{1-1.5\omega_D}\right]   
\end{equation}
where $\omega_D=0.058$ accounts for the D-state admixture in the deuteron wave function.\cite{wd} \\

From Figure 1,2 and 3, we observe that our analytical models represented by Eq.(\ref{eqn:solnspol2})(MOC2) and Eq.(\ref{solnlagmethod1})(LM) are in good agreement with the experimental data upto the $x$ value $x\leq0.4$. But our solution given by method of characteristics, Eq.(\ref{eqn:solnspol1})(MOC1) overshoots the data beyond $x\geq 0.28$. Thus we can conclude that the valid range of $x$ for the three analytical models is $x\leq0.28$, above which the small $x$ approximation is no more valid.\\

Further we test the compatibility of the three analytical solutions with the $Q^2$ evolution within the valid small $x$ range of the experimental data (both HERMES and COMPASS). For HERMES, small $x$ data $x\leq0.28$ is within the $Q^2$ range $Q^2\leq 6.94 GeV^2$, whereas for COMPASS it is within the $Q^2$ range, $Q^2\leq 17.2 GeV^2$. We therefore confine our comparison within this $Q^2$ range $1.1 GeV^2\leq Q^2\leq 17.2GeV^2$.\\

Figure 4,5 and 6 shows the compatibility of our analytical models with COMPASS data and Figure 7, 8 and 9 shows the same for HERMES data separately. From the figures we observe that our analytical models are reasonably cosistent within the experimentally accessible small $x$ range of data $0.0046\leq x\leq0.28$ and $Q^2$ range $1.1GeV^2\leq Q^2 \leq17.2 GeV^2$.\\

\begin{figure}[hb]
\begin{center}
\includegraphics[width=4in]{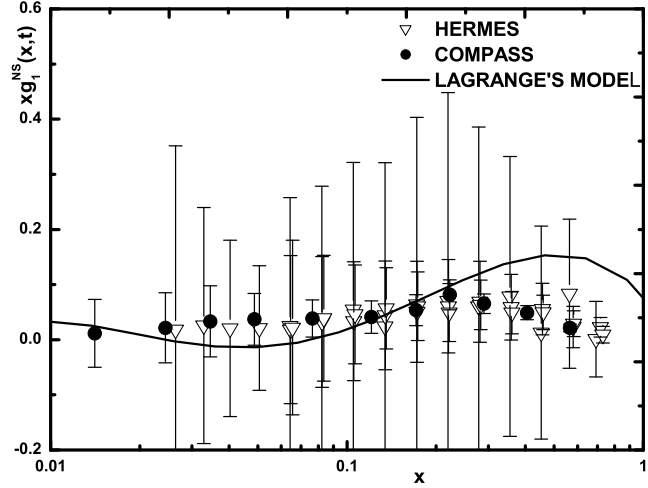}
\end{center}
\vspace{-0.10in}
\caption[Polarized non-singlet  structure function $xg_1^{NS}(x,t)$]{Polarized non-singlet  structure function $g_1^{NS}(x,t)$ as function of $x$ at different $Q^2$ according to Eq.(\ref{solnlagmethod1}). Data from refs \cite{HERMES} and \cite{compass}}
\label{fig:1}
\end{figure}

\begin{figure}
\begin{center}
\includegraphics[width=4in]{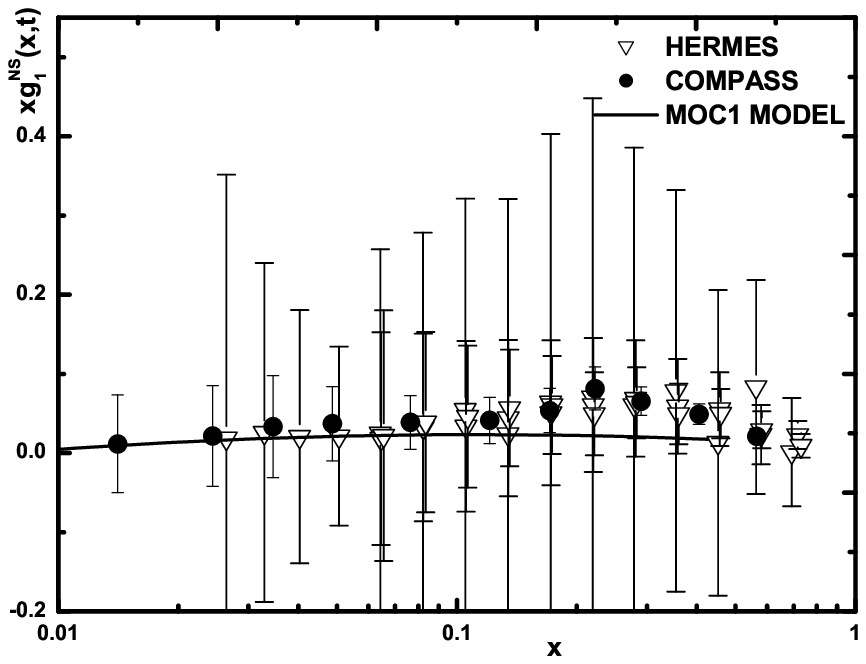}
\end{center}
\vspace{-0.10in}
\caption[Polarized non-singlet  structure function $xg_1^{NS}(x,t)$]{Polarized non-singlet  structure function $g_1^{NS}(x,t)$ as function of $x$ at different $Q^2$ according to Eq.(\ref{eqn:solnspol1}). Data from refs \cite{HERMES} and \cite{compass} }
\label{fig:2}
\end{figure}

\begin{figure}
\begin{center}
\includegraphics[width=4in]{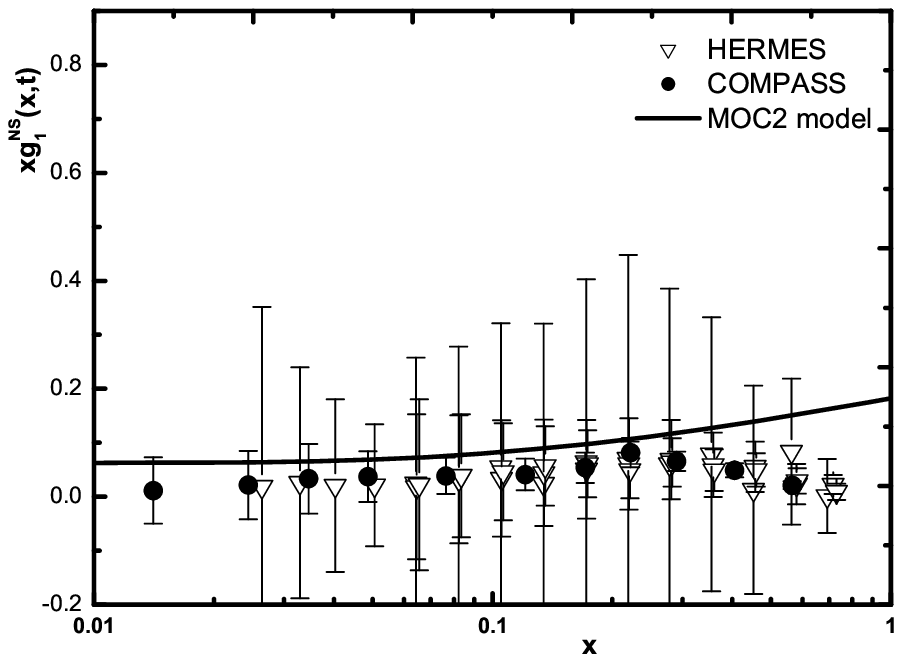}
\end{center}
\vspace{-0.10in}
\caption[Polarized non-singlet  structure function $xg_1^{NS}(x,t)$]{Polarized non-singlet  structure function $g_1^{NS}(x,t)$ as function of $x$ at different $Q^2$ according to Eq.(\ref{eqn:solnspol2}). Data from refs \cite{HERMES} and \cite{compass}}
\label{fig:3}
\end{figure}

\begin{figure}
\begin{center}
\includegraphics[width=4in]{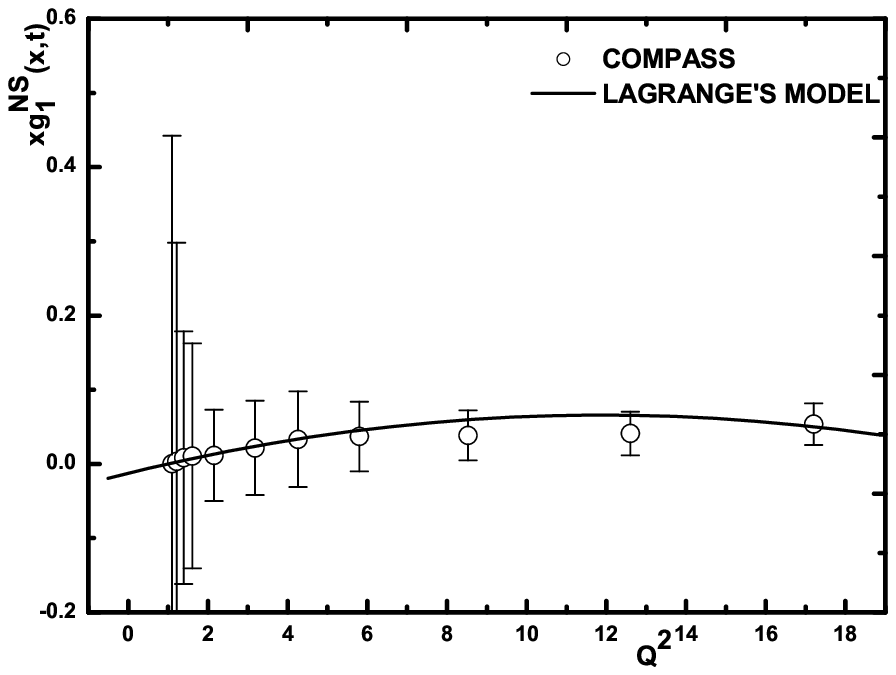}
\end{center}
\vspace{-0.10in}
\caption[Polarized non-singlet  structure function $g_1^{NS}(x,t)$]{Polarized non-singlet  structure function $g_1^{NS}(x,t)$ as function of $Q^2$ at different $x$ according to Eq.(\ref{solnlagmethod1}). Data from refs  \cite{compass}}
\label{fig:4}
\end{figure}

\begin{figure}
\begin{center}
\includegraphics[width=4in]{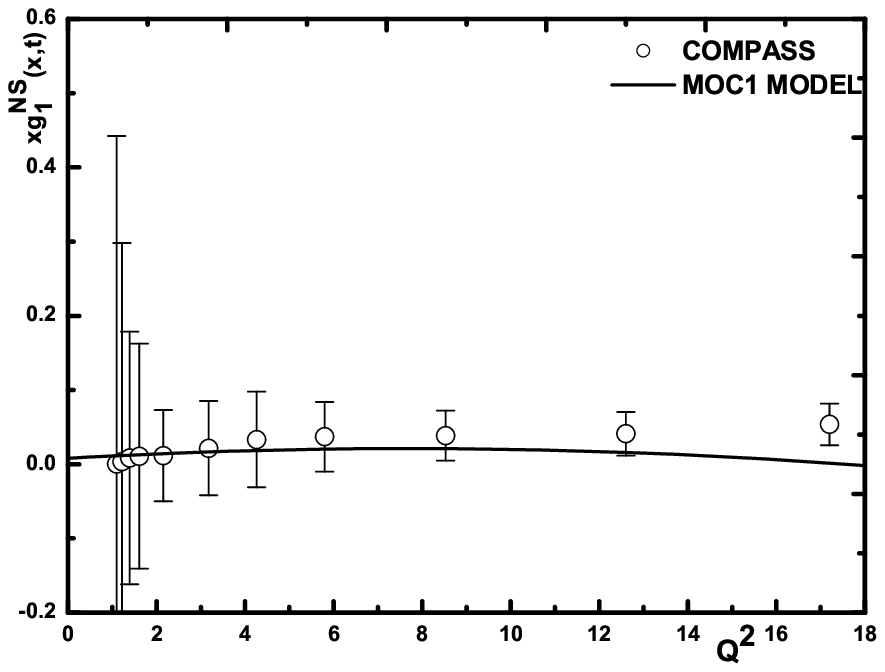}
\end{center}
\vspace{-0.10in}
\caption[Polarized non-singlet  structure function $g_1^{NS}(x,t)$]{Polarized non-singlet  structure function $g_1^{NS}(x,t)$ as function of $Q^2$ at different $x$ according to Eq.(\ref{eqn:solnspol1}). Data from refs \cite{compass}}
\label{fig:5}
\end{figure}

\begin{figure}
\begin{center}
\includegraphics[width=4in]{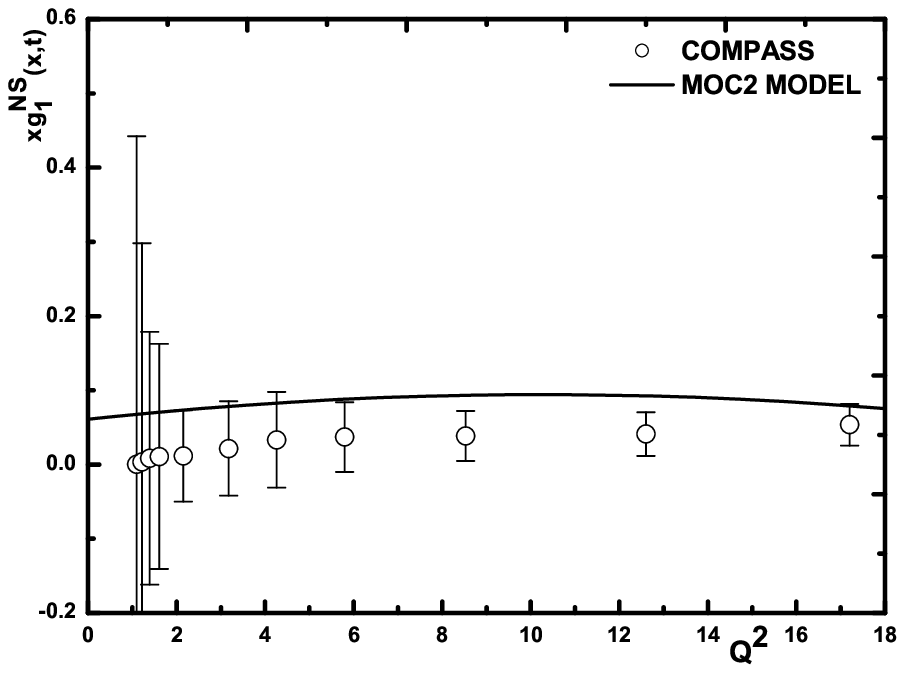}
\end{center}
\vspace{-0.10in}
\caption[Polarized non-singlet  structure function $g_1^{NS}(x,t)$]{Polarized non-singlet  structure function $g_1^{NS}(x,t)$ as function of $Q^2$ at different $x$ according to Eq.(\ref{eqn:solnspol2}). Data from refs \cite{compass}}
\label{fig:6}
\end{figure}

\begin{figure}
\begin{center}
\includegraphics[width=4in]{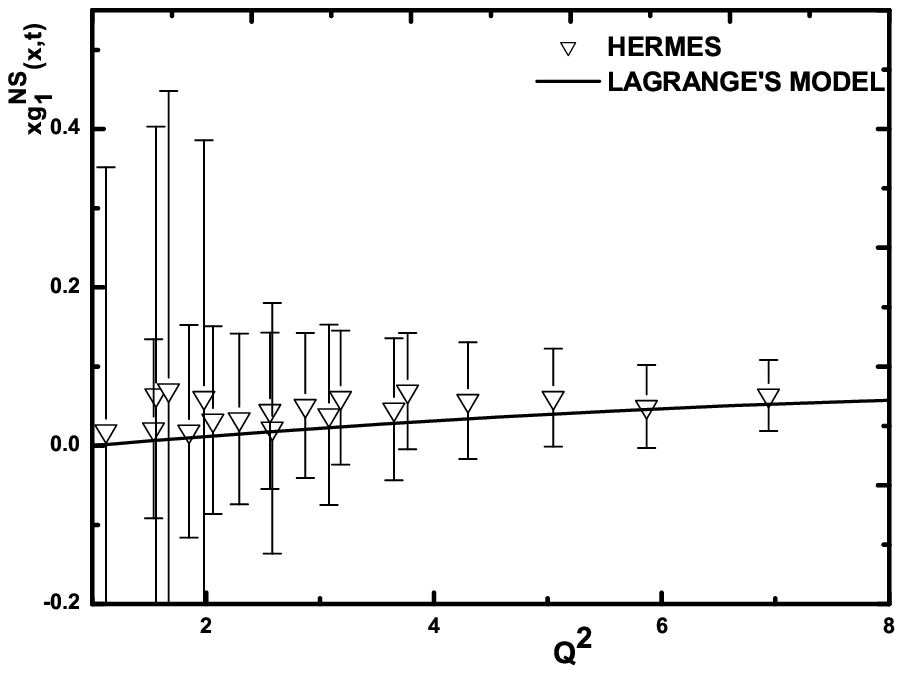}
\end{center}
\vspace{-0.10in}
\caption[Polarized non-singlet  structure function $g_1^{NS}(x,t)$]{Polarized non-singlet  structure function $g_1^{NS}(x,t)$ as function of $Q^2$ at different $x$ according to Eq.(\ref{solnlagmethod1}). Data from refs  \cite{HERMES}}
\label{fig:4}
\end{figure}

\begin{figure}
\begin{center}
\includegraphics[width=4in]{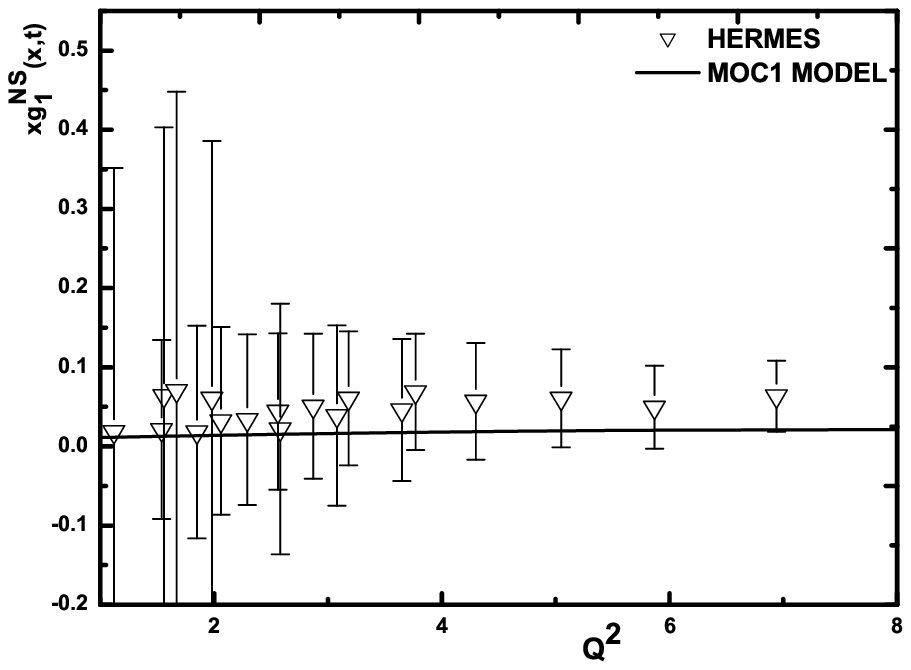}
\end{center}
\vspace{-0.10in}
\caption[Polarized non-singlet  structure function $g_1^{NS}(x,t)$]{Polarized non-singlet  structure function $g_1^{NS}(x,t)$ as function of $Q^2$ at different $x$ according to Eq.(\ref{eqn:solnspol1}). Data from refs \cite{HERMES}}
\label{fig:5}
\end{figure}

\begin{figure}
\begin{center}
\includegraphics[width=4in]{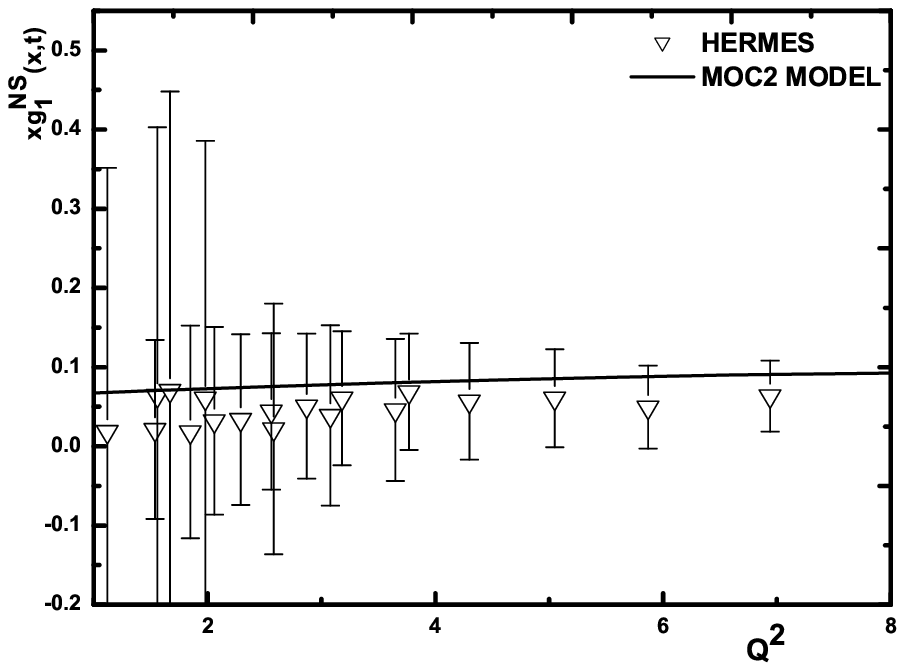}
\end{center}
\vspace{-0.10in}
\caption[Polarized non-singlet  structure function $g_1^{NS}(x,t)$]{Polarized non-singlet  structure function $g_1^{NS}(x,t)$ as function of $Q^2$ at different $x$ according to Eq.(\ref{eqn:solnspol2}). Data from refs \cite{HERMES}}
\label{fig:6}
\end{figure}

%\begin{figure}
%\begin{center}
%\includegraphics[width=4in]{last_moment.EPS}
%\end{center}
%\vspace{-0.10in}
%\caption[Polarized non-singlet  structure function $g_1^{NS}(x,t)$]{First moments of polarized non-singlet  structure function $g_1^{NS}(x,t)$ as function of $x_{min}$ according to Eq.(\ref{eqn:solnspol1}) and Eq.(\ref{eqn:solnspol2})Eq.(\ref{solnlagmethod1}) }
%\label{fig:7}
%\end{figure}

We perform a $ \chi^2$ test  using the formula $\chi^2 =\Sigma_i \frac{(X_{th}-X_{ex})^2}{\sigma^2 } $, to check the quantitative estimate of the goodness of fit among the solutions obtained by two analytical methods with the experimental data. In Table 1 we give the $ \chi^2/d.o.f$ for each analytical solutions given by Eq.(\ref{eqn:solnspol1}), Eq.(\ref{eqn:solnspol2}), Eq.(\ref{solnlagmethod1}). The d.o.f. for HERMES and COMPASS are 36 and 14 respectively.\\
\begin{table}[!ht]
\label{table:chisquare}
\begin{center}
\caption[$\chi^2/d.o.f.$ distribution]{$\chi^2/d.o.f.$ values for Eq.(\ref{eqn:solnspol1}), Eq.(\ref{eqn:solnspol2}) and Eq.(\ref{solnlagmethod1})}
\vspace{0.1in}
\begin{tabular}{|l|l|c|c|}
\hline
Method & Solutions & HERMES Collaboration & COMPASS Collaboration \\ 
\hline
Method of characteristics & MOC1,Eq.(\ref{eqn:solnspol1}) & 0.0218 & 0.57 \\
\hline
Method of characteristics & MOC2,Eq.(\ref{eqn:solnspol2}) & 0.055 & 0.58\\
\hline 
Lagrange's Method & Eq.(\ref{solnlagmethod1}) & 0.054 & 0.616\\ 
\hline
\end{tabular}
\end{center}
\end{table}

From the $\chi^2$ analysis as given in Table 1, we infer that the analytical solution given by Eq.(\ref{eqn:solnspol1})(MOC1) fares better that the other two Eq.(\ref{eqn:solnspol2}) and Eq.(\ref{solnlagmethod1}).\\

Let us now comment on the Bjorken sum rule \cite{Bjorken1} in the context of present work. As is well known, important informations about the spin structure of nucleon can be extracted from the first moment of spin structure function $g_1^{NS}$, known as Bjorken sum rule. The Bjorken integral defined as,
\begin{equation}
\Gamma_1^{NS}=\int_0^{1}(g_1^{P}(x,Q^2)-g_1^{n}(x,Q^2))dx\equiv \int_0^{1}g_1^{NS}(x,Q^2)dx
\end{equation} 
Where $g_1^{NS}(x,Q^2)$ is expressed in Eq.(\ref{g1ns}) in terms of $\Delta q^{NS}(x,Q^2)$.\\
To evaluate theoretically $\Gamma_1^{NS}$, one needs information about $\Delta q^{NS}(x,Q^2)$ in the entire physical region of $x$, $(0\leq x\leq 1)$. A model whose validity is tested only in a limited $x$ range $(x_{a}\leq x\leq x_{b})$, one can obtain only partial information about $\Gamma_1^{NS}$. i.e.
\begin{equation}
\hat\Gamma_1^{NS}=\int_{x_a}^{x_b}g_1^{NS}(x,Q^2)dx
\end{equation}
which gives information about the contribution to the $\Gamma_1^{NS}$ from the partons having fractional momentum $x_{a}\leq x\leq x_{b}$, which should invariably be less than its exact experimental measured value, i.e. $\Gamma_1^{NS}\gg  \hat\Gamma_1^{NS}$. A similar analysis for partial momentum sum rule \cite{akbari} has been reported recently. It is to be noted that in ref \cite{12}, $\Gamma_1^{NS}$ was calculated itself as $\hat\Gamma_1^{NS}$, even though our model was valid for $x$ range $0.02\leq x\leq 0.3$ only. The E155 collaboration at SLAC found $\Gamma_1^{NS}=0.176\pm0.003\pm0.007$ at $Q^2=5 GeV^2$, which is confirmative with SMC results, $\Gamma_1^{NS}=0.174\pm0.024\pm0.002$ at $Q^2=5 GeV^2$. For the HERMES collaboration \cite{HERMES}, the cumulative range of $\Gamma_1^{NS}$ within the range $(0.21\leq x\leq 0.9)$ at $Q^2=2.5 GeV^2$ is found to be, $\Gamma_1^{NS}=0.1477\pm0.0055\pm0.01102$. In recent analysis for COMPASS \cite{COMPASS}, the $\Gamma_1^{NS}$ is evaluated in the range $0.004\leq x \leq 0.7$, to be $\Gamma_1^{NS}=0.175\pm 0.009\pm 0.015$ at $Q^2=3GeV^2$. COMPASS has further measured $\hat\Gamma_1^{NS}$ separately as shown in Table 2 below, where $\Gamma_1^{NS}=0.190 \pm 0.009 \pm 0.015$. Let us now estimate the partial value $\hat\Gamma_1^{NS}$ and compare its contribution to the measured value of $\Gamma_1^{NS}$, for the analytical models.\\

\begin{table}[!ht]
\label{table:Integrals of $g_1^{NS}$}
\begin{center}
\caption[Partial Integrals of $g_1^{NS}$]{First moment $\Gamma_1^{NS}$ at $Q^2=3GeV^2$ from the COMPASS\cite{COMPASS} data points.}
\vspace{0.1in}
\begin{tabular}{|c|c|}
\hline
$x$ range & $\Gamma_1^{NS}$\\ 
\hline
0 - 0.004 & 0.0098\\
\hline
0.004 - 0.7 & $0.175 \pm 0.009 \pm 0.015$\\
\hline 
0.7 - 1.0  & 0.0048 \\ 
\hline
0 - 1 & $0.190 \pm 0.009 \pm 0.015$ \\
\hline
\end{tabular}
\end{center}
\end{table}

In Table 3, we show a prediction to partial $\hat\Gamma_1^{NS}$, for $Q^2=2.5 GeV^2$ and $5GeV^2$, taking contribution from the region $0.0046\leq x\leq0.28$, where the analytical models work reasonably. The results are far less than the corresponding experimental values of $\Gamma_1^{NS}$. The remaining part of $\Gamma_1^{NS}$ is contributed by the small $x$ and large $x$ partons having $x\leq 0.0046$ and $x\geq 0.28$ respectively. \\

\begin{table}[!ht]
\label{table:Integrals of $g_1^{NS}$}
\begin{center}
\caption[Partial Integrals of $g_1^{NS}$]{Predictions for the partial integrals of $g_1^{NS}$ ($\hat\Gamma_1^{NS}$) obtained for the solutions Eq.(\ref{eqn:solnspol1}), Eq.(\ref{eqn:solnspol2}) and Eq.(\ref{solnlagmethod1}) with $x_a=0.0046$ and $x_b=0.28$ for $Q^2=2.5 GeV^2$ and $5GeV^2$.}
\vspace{0.1in}
\begin{tabular}{|c|c|c|}
\hline
 Solutions &$Q^2=2.5GeV^2$ & $Q^2=5GeV^2$\\ 
\hline
 MOC1,Eq.(\ref{eqn:solnspol1}) & 0.08939 & 0.08532 \\
\hline
MOC2,Eq.(\ref{eqn:solnspol2}) & 0.03579 & 0.03426\\
\hline 
Lagrange's method Eq.(\ref{solnlagmethod1})  & 0.054 & 0.0667\\ 
\hline
\end{tabular}
\end{center}
\end{table}
As the present formalism is expected to be valid for small $x$, we also estimate the prediction for $\hat\Gamma_1^{NS}$, for the present models, assuming its validity in the small $x$ range $0\leq x\leq 0.0046$ separately. The results are shown in Table 4 for $Q^2=2.5 GeV^2$ and $5GeV^2$, which below the experimental value \cite{COMPASS}, suggests the large $x$,($x\geq 0.28$) contribution is necessary in case of our analytical models to account for the experimental value.  \\ 

\begin{table}[!ht]
\label{table:Integrals of $g_1^{NS}$}
\begin{center}
\caption[Integrals of $g_1^{NS}$]{Integrals of $g_1^{NS}$ for analytical solutions Eq.(\ref{eqn:solnspol1}), Eq.(\ref{eqn:solnspol2}), Eq.(\ref{solnlagmethod1}) in the limited small $x$ range $0\leq x\leq 0.0046$ at $Q^2=2.5GeV^2$ and $5GeV^2$. }
\vspace{0.1in}
\begin{tabular}{|l|c|c|}
\hline
  Solutions &$Q^2=2.5 GeV^2$ & $Q^2=5 GeV^2$ \\ 
\hline
 MOC1,Eq.(\ref{eqn:solnspol1}) & 0.0008853 & 0.0009036 \\
\hline
 MOC2,Eq.(\ref{eqn:solnspol2}) &0.0004845 & 0.0004753\\
\hline 
Lagrange's method,Eq.(\ref{solnlagmethod1}) & 0.0001248 & 0.0001529\\ 
\hline
\end{tabular}
\end{center}
\end{table}
Though our analytical models works reasonably well within a small $(x,Q^2)$ range as discussed above, we still check their contribution to $\Gamma_1^{NS}$ in the high $x$ range $(0.0046\leq x \leq0.7)$, for completeness of the comparative study. The respective values of $\Gamma_1^{NS}$ at $Q^2=2.5GeV^2$ and $5GeV^2$ are shown in Table 5 below. The values are lower than the experimental value for COMPASS, $0.175\pm 0.009\pm 0.015$ within the region $0.004\leq x \leq0.7$ at $Q^2=3GeV^2$. But at increased $Q^2$ our analytical solution derived by Lagrange's method give higher values. Hewever as our analytical solution Eq.(\ref{eqn:solnspol1}) given by method of characteristics is not valid beyond $x\geq 0.28$, hence we do not get a value of $\Gamma_1^{NS}$ for that model for high $x$ range.\\
\begin{table}[!ht]
\label{table:Integrals of $g_1^{NS}$}
\begin{center}
\caption[Integrals of $g_1^{NS}$]{Integrals of $g_1^{NS}$ for analytical solutions Eq.(\ref{eqn:solnspol2}), Eq.(\ref{solnlagmethod1}) in the high $x$ range $0.0046\leq x\leq 0.7$ at $Q^2=2.5GeV^2$ and $5GeV^2$ .}
\vspace{0.1in}
\begin{tabular}{|l|c|c|}
\hline
  Solutions &$Q^2=2.5 GeV^2$ & $Q^2=5 GeV^2$ \\ 
\hline
 MOC2,Eq.(\ref{eqn:solnspol2}) &0.131 & 0.124\\
\hline 
Lagrange's method,Eq.(\ref{solnlagmethod1}) & 0.161 & 0.197\\ 
\hline
\end{tabular}
\end{center}
\end{table}

%For completeness in Figure 8, we show the graphical representation of the exponents of ($\frac{t}{t_0}$) as defined in Eq.(\ref{eqn:solnspol1}), Eq.(\ref{eqn:solnspol2}) and Eq.(\ref{solnlagmethod1}) for small $x$. It indicates that the exponents of the solutions of Lagrange's method are invariably positive definite, as well as the corresponding exponents for Method of Chracteristics.\\

\section{Conclusion}
\label{sec:conclusion} 
In this work we have calculated the non-singlet spin structure function $g_1^{NS}$ using two analytical methods: Lagrange's and Method of Characteristics. The analytical solutions are in good agreement with the experimental data from both HERMES and COMPASS within a comparatively small $x$ range $0.0046\leq x\leq 0.28$ with $Q^2$ range $1.1 GeV^2 \leq Q^2 \leq 17.2 GeV^2$. From our analysis (both graphical and $\chi^2$) we conclude that our analytical solution Eq.(\ref{eqn:solnspol1}) obtained by method of characteristics compares best in the $(x,Q^2)$ range defined above. We have also calculated the partial momentum fractions, carried by small $x$ ($0.0046\leq x\leq 0.28$) non-singlet partons for the analytical models. The behaviour of analytical models in NLO is currently under study.

\end{document}